\documentclass[esprc2,12pt]{article}
\usepackage{graphicx}
\usepackage[top=3cm, bottom=2cm, left=3cm, right=3cm]{geometry}

\title{The role of the Galactic Halo and the Single Source in the formation of the 
cosmic ray anisotropy} 
\author{$^{*}$A.D. Erlykin $^{1,2}$ and A.W. Wolfendale $^{2}$\\
$(1)$ P N Lebedev Physical Institute, Moscow, Russia.\\
$(2)$ Physics Department, Durham University,\\ Durham, DH1 3LE, UK}

\begin{document}
\maketitle

\footnote{$^{*}$Corresponding author: tel +74991358737 \\ 
 E-mail address: erlykin@sci.lebedev.ru}

\begin{abstract}
The existence of the cosmic ray Halo in our Galaxy has been discussed for more than 
half a century. If it is real it could help to explain some puzzling features 
of the cosmic ray flux: its small radial gradient, nearly perfect isotropy and the low 
level of the fine structure in the energy spectra of the various particles. All these 
features could be understood if: (a) the Halo has a big size (b) cosmic rays in the 
Halo have a unform spatial or radial distribution and (c) the cosmic ray density in the
 Halo is comparable or even higher than that in the Galactic Disk. The main topic of 
the paper concerns the present status of the anisotropy and a model for its formation. 
In our model the extremely small amplitude of the dipole anisotropy is due to the 
dilution of the anisotropy in the Disk by the dominating isotropic cosmic rays from the
 Halo. Some minor deviations from complete isotropy in the sub-PeV and PeV energy 
regions point out to the possible contribution of the Single Source with the phase of 
its first harmonic opposite to the 
phase produced by the Disk. 
\end{abstract}


\section{Introduction}

The observed cosmic rays (CR) have several puzzling features which need to be explained
 and these are now listed: 

(a) Firstly, there is only a small radial gradient in the Galactic Disk (GD) in
 contrast with expectation, the reason is as follows. The most viable theory of the CR 
origin is that they are generated in the supernova (SN) explosions and 
acclerated by the shock waves in the supernova remnants (SNR) \cite{Ginz1}. According 
to recent studies \cite{Case,Green} the Galactocentric radial distribution of SNR is 
such that they are mostly concentrated in the Inner Galaxy with the maximum GD surface 
density at a Galactocentric radius of about $R\approx 3-4$ kpc followed by a rapid 
decrease at larger $R$. According to the model calculations \cite{EW1} the radial CR 
gradient coincides with that of SNR and at the Sun, of radial distance of 8.3 kpc, it 
 should be equal to $S = d(lnI)/dR = (-0.17\pm 0.05)$ kpc$^{-1}$. Here $I$ is the CR 
intensity or the SNR density. However, the experimental values for the Outer Galaxy 
derived by us from the gamma-ray emissivity profile are $(-0.05\pm 0.03)$ kpc$^{-1}$ 
for both the second and third quadrants \cite{Abdo,Acker}, so that the observed CR 
radial distribution is significantly flatter than that expected from the distribution 
of their proposed sources.

(b) Secondly, there is the surprising near-isotropy of the CR arrival directions. In 
the sub-PeV region the CR intensity is relatively high and allows the collection of 
good statistics with detectors of a reasonable size during an acceptable time. 
Theoretical calculations predict a the slow rise for the amplitude {\em A} of the first
 harmonic with energy {\em E} as $A \sim E^{0.3-0.5}$. On the opposite side the 
experimental measurements indicate a decreasing amplitude above a few TeV with a 
minimum of about $A\sim 2\cdot 10^{-4}$ for $E\sim (0.1-0.3)$PeV \cite{Guill,EW2}. The 
attempt to explain this decrease by an accidental spatial configuration of the sources 
is difficult because it shows that the probability of such a favorable configuration is
 definitely lower than a few percent \cite{EW3}. 

(c) The third puzzle is connected with the observed shape of the CR energy spectrum. 
Due to the stochastic distribution of the SNR in space and time there should be fine 
structure in the spectrum at some level. All realistic simulations confirm the possible
 existence of such structures \cite{EW1}. However, so far, only two structures are 
firmly established: the so called 'knee' at 3-4 PeV and the 'ankle' at 3-4 EeV. Below 
the knee, and between the knee and the ankle, measurements show quite a regular power 
law shape of the spectrum with only minor structure.

In the last decade due to improvements in the energy resolution and increased 
statistics several works find hints of fine structure both below \cite{ATICn,CREAM,
PAMELAn} and above the knee \cite{KASCADE1,GAMMA,TUNKA,IceTop,Yakutsk1}. In the region 
below the knee the experiments indicate a possible flattening of the proton and nuclei 
spectra above a rigidity of 200 GV. A similar flattening was also found in the primary 
electron plus positron spectrum \cite{ATICe1,ATICe2,PAMELAe}. However, the latest 
precise data from AMS-02 experiment do not show such features, except for positrons, in
 the same energy region below the knee \cite{AMS2}. The clarification of the situation 
is the duty of the experimental groups, but in any case the discussed irregularities,
 if they exist, are relatively small and do not disprove the basic feature of the CR 
energy spectrum, viz its nearly perfect regular power law shape.   
               
The present paper is an attempt to find a reasonable explanation of this puzzle, with 
special attention given to the large-scale anisotropy.

\section{The present status of the cosmic-ray large-scale anisotropy}
The large-scale anisotropy of CR is usually described by the  amplitude and phase of 
the first and second harmonics. The phase, expressed in terms of the right ascension 
($RA$), is the direction of the maximum CR intensity. Due to the extremely small 
deviations from isotropy the measurement of these deviations requires large statistics.
Until recently they were concentrated mostly in the TeV and sub-PeV energy regions. 
Only in the last decade have arrays with the necessary large aperture: Pierre Auger
 Observatory, Telescope Array, IceCube and IceTop, Yakutsk and KASCADE-Grande 
accumulated the good statistics necessary to probe PeV and even EeV energies. Also, 
large EAS arrays such as ARGO-YBJ and Tibet III have produced precise results in the 
sub-PeV region. Figure 1 shows the present situation with the measurements of the 
dipole anisotropy.
\begin{figure}[hpt]
\begin{center}
\includegraphics[height=15cm,width=12cm,angle=-90]{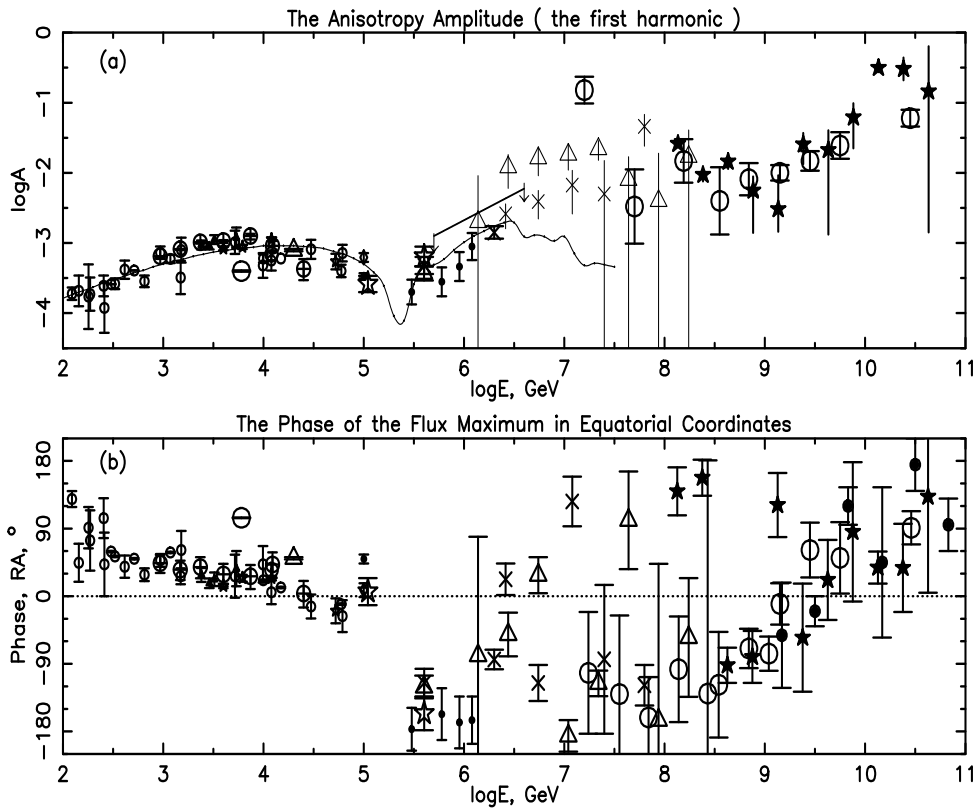}
\end{center}
\caption{\footnotesize Amplitude (a) and Phase (b) of the first harmonic of the CR 
anisotropy. The data at sub-PeV energies have been reproduced from \cite{EW2}. At 
higher energies they were taken from the Pierre Auger Observatory \cite{PAO}: 
($\bigcirc$), Telescope Array ( only phases were published ) \cite{TA} ($\bullet$), 
Yakutsk \cite{TA,Yakutsk2} ($\star$), KASCADE-Grande \cite{KASCADE2} ($\times$), Akeno 
\cite{Akeno} ($\bigtriangleup$). The full line in (a) relates to our model and is 
reproduced here from Figure 2b (see \S5).}   
\label{fig:fig1}
\end{figure}   

In general there are several features noticable in this survey: \\
(i) a good consistency of the results at energies up to a few PeV; \\
(ii) the extremely small $\sim 10^{-4} \div 10^{-3}$ amplitude of the anisotropy; \\
(iii) the visible rise of the amplitude $A$ with energy $E$ up to $logE,GeV 
\approx 4$; \\
(iv) a moderate fall of the amplitude above $logE \approx 4$ up to a minimum at 
$logE,GeV \approx 5.3 \div 5.5$; \\
(v) the rise of the amplitude beyond this minimum up to a few PeV; \\ 
(vi) the approximately constant phase at low energies which suddenly changes its 
direction at about the same energy of $logE \approx 5.3 \div 5.5$ 
where the amplitude has a minimum; \\
(vii) in the PeV region, where the rise of the amplitude is observed, the phase has 
an apparent trend to recover up to its previous direction close to $RA \sim 0$.   

In what follows we shall endeavour to build a model which can reproduce these features 
with the minimum number of assumptions. This model contains three basic ingredients:
the Galactic Disk, the Halo and the Single Source (SS). Although we separate here the 
role of the Single Source, we understand that, in fact, it is just part of CR in 
the Disk.
\section{Interrelation between the Galactic Disk and the Halo}
Stars in our Milky Way Galaxy are concentrated in the Disk and likewise are SNR. For 
simplicity we consider SNR as the only source of CR, their acceleration and the energy.
 As was already mentioned, the Galactocentric radial distribution of SNR is 
non-uniform.

It is clear that to fit the observations ( see puzzle (a) in the Introduction ) we have
 either to abandon the model with SNR as the dominant CR sources, because it inevitably
 gives a non-uniform spatial CR distribution, or to add a CR component with a uniform 
spatial, isotropic directional and uniform temporal distribution. The latter 
requirement is not strict, but the uniform temporal distribution helps to make the 
spatial distribution uniform, since stochastic explosions of SN in different parts of 
the Galaxy make the spatial distribution patchy. Another requirement is that to have a 
visible effect the CR energy density of this additional component should be comparable 
with that observed: $\sim 10^{-12}erg/cm^3$.

Extragalactic CR cannot be such an additional component since if CR are supplied by 
galaxies similar to our Milky Way, with a power $\sim10^{48}erg/year$ and the 
volume density of galaxies in the extragalactic space is $6.4\cdot10^{-2}Mpc^{-3}$ then
 even the whole age of the Universe of $14\cdot10^9 year$ is not enough to fill its 
volume with the required density. The most likely source of the additional component is
 the Galactic Halo with its role as a mediator between the Disk and Extragalactic 
Space.
\subsection{The existence of the Halo}
The existence and the possible role of our Milky Way Halo has been discussed for 8 
decades. One of the most comprehensive reviews was made by V.L.Ginzburg in the 
seventies \cite{Ginz2}. Since that time considerable progress in the studies of our 
Galaxy and its environment have been made. Radiohaloes have been found in several 
spiral galaxies seen edge-on and similar to our Milky Way (eg NGC 4631, NGC 891). Big 
'fountains' of matter emerging from our Galaxy were observed in radio - such as the 
North Polar Spur. 
X-ray observations with the ROSAT, XMM-Newton, Suzaku and Chandra satellites have found
 evidence for the emission of hot gas far above the Galactic Disk with features which 
favour its origin from the violent processes in the Disk. Also observed have been high 
and uniform temperature of the order $(2-3)\cdot 10^6K^\circ$, an increase of emission 
towards the Galactic Center, patchiness of the gas density, a Galactic latitude 
dependence which does not fit the disk-plane parallel model (see the bibliography in 
\cite{XMM}). Gamma ray studies of the synchrotron emission favour a vertical scale for 
the gamma-ray intensity of about 10 kpc, which is already comparable with the Disk 
radius of 15 kpc \cite{Strong}. They have also revealed the existence of big bubbles of
 emitting hot gas above the Disk emerging from the region close to the Galactic Center,
 the so called 'Fermi Bubbles' \cite{FERMI}.
 
However, all the above indications cannot be accepted as proof of the existence of
a hadronic CR Halo fed by violent processes in the Disk. As an alternative 
explanation the hot gas could be supplied by the accretion of the extragalactic (EG) 
gas by the Disk. Radio and soft gamma-ray emission are definitely connected with the 
synchrotron radiation of relativistic electrons in the Galactic magnetic fields, but 
electrons comprise only a minor part $\sim1$\% of all CR. The most convincing 
indication of the 
existence of the hadronic Halo is still measurements of radioactive isotopes in CR, 
which together with other arguments are nevertheless indirect. 
 However, as E.L.Feinberg once said, 'the proof in cosmic 
rays is of a special kind - it is the aggregate of indications'. Thus we consider the 
above aggregate seriously and shortly examine a model with a hadronic Halo. 
\subsection{The existence of the Single Source}
The existence of the Single Source has been proposed by us to explain the puzzling 
sharpness of the knee in the EAS size spectrum ( see \cite{EW4} and later publications 
). The physical basis of this proposal is the evident non-uniformity of the spatial and
 temporal distributions of SN explosions and subsequent SNR. As a result one SN could 
explode not very long ago and close to the solar system.. Its contribution to the CR 
intensity is rather high and it gives rise to a small peak (~knee~) above the 
background from other SNR - it is our Single Source. 

In the last decade, the 
non-uniformity of the CR source distribution has been actively discussed in connection 
with the discovery of the peculiarities of the primary electron and positron spectra  
\cite{Panov}. Due to the limited range and the lifetime of electrons and positrons, 
their emitting sources should be nearby and relatively young like the Single Source 
responsible for the knee. It does not mean that they are identical, but if the 
observed peculiarities are real they stress the importance of the non-uniformity of the
 CR source distribution in space and time. We shall also consider the Single Source
as a real object and examine the consequences of its existence.  
\section{The model}
\subsection{The Galactic Disk and the Halo}
The model is actually the generalisation of the Leaky Box scenario for the composite 
system of Disk+Halo. If a SNR converts a part of its kinetic energy into CR with a 
power law energy spectrum, then the expression for this spectrum is:
\begin{equation}
I(E)=\frac{W(\gamma-2)c}{4\pi V E_{min}^2}(\frac{E}{E_{min}})^{-\gamma}
\end{equation} 
Here, $W$ is the total CR energy, $\gamma$ is the slope index of the differential power
 law spectrum, $c$ is the speed of light, $V$ is the volume containing the CR, 
$E_{min}$ is the minimum 
energy of the spectrum. In the following calculations we adopt $W = 10^{50}$erg, 
$\gamma = 2.15$ from our model of SNR acceleration \cite{EW6} and $E_{min}$ = 1GeV. 
We assume that the CR distribution in the volume $V$ is uniform and isotropic. The 
maximum energy of the spectrum is not important if $E_{max}\gg E_{min}$.

If the system of Disk+Halo is in dynamic equilibrium the amount of CR energy 
supplied by SNR in the Disk, with rate $\nu$, should be equal to the CR energy 
supplied by the Disk to the Halo and eventually transferred from the Halo to EG space. 
The energy spectrum of CR residing permanently in the Disk and the Halo depends on 
their lifetime $\tau (E)$. If this lifetime is also dependent on energy as
$\tau = \tau_0 (\frac{E}{E_{min}})^{-\delta}$, then the energy spectrum of CR in the 
Disk and the Halo is:
\begin{equation}
I(E)=\frac{\nu \tau_0 W(\gamma-2)c}{4\pi V E_{min}^2}(\frac{E}{E_{min}})^
{-(\gamma+\delta)}
\end{equation}
Apparently, $\tau_0, V$ and $\delta$ are different for the Disk and the Halo and should 
be indexed as $\tau_{0,d,h}$, $ V_{d,h}$ and $\delta_{d,h}$.
   
It is known from the measurements of the secondary to primary CR ratio and of 
radioactive 
isotopes that CR spend most of their lifetime in the Halo \cite{Connell}. Averaging
their values for different isotopes we assume that $\tau_{0,h} = 1.5\cdot 10^7$year and
 $\delta_h = 0.5$ \cite{EW6}. We consider the Halo as an ellipsoid with radius 
$R_h = 15kpc$ and height $H_h = 10 kpc$, so that its volume is equal to 
$V_h = \frac{4}{3} H_h R_h^2 = 3000 kpc^3$. For the rate of SN explosions we adopt 
$\nu = 10^{-2} year^{-1}$. Calculations performed with these numerical
 parameters give CR intensity significantly less than the experimental values. However,
 taking into account all the assumptions and uncertainties of the parameters used in 
the expression (2) we can certainly make the calculated results closer to the 
experimental values. The most doubtful assumption is the requirement of a uniform 
spatial distribution inside the Halo, the shape and the size of the Halo and $E_{min}$.

The need for a nearly uniform intensity in the Halo suggests some form of reflecting 
walls. There are various possibilities, including our Galaxy being in a Giant Bubble. 
As is well known, there are Bubbles of a variety of sizes round stars, both singly and 
in clusters and an extrapolation of the idea of stellar winds can be extended to 
Galactic Winds. Another possibility is that there is shear-induced turbulence caused by
 the rotation of our galaxy with respect to the nearby Intergalactic Medium. Since all 
these features and possible non-uniformity of the CR distribution in the Halo are 
unknown we use the simplest scenario assuming the uniform spatial CR distribution in 
the Halo.    
 
 Finally, we normalise the absolute
intensity of calculated spectrum to the results of experimental measurements of 
the all-particle spectrum. It is shown in Figure 2a and denoted as 'HALO'.       

\begin{figure}[ht]
\begin{center}
\includegraphics[height=15cm,width=12cm,angle=-90]{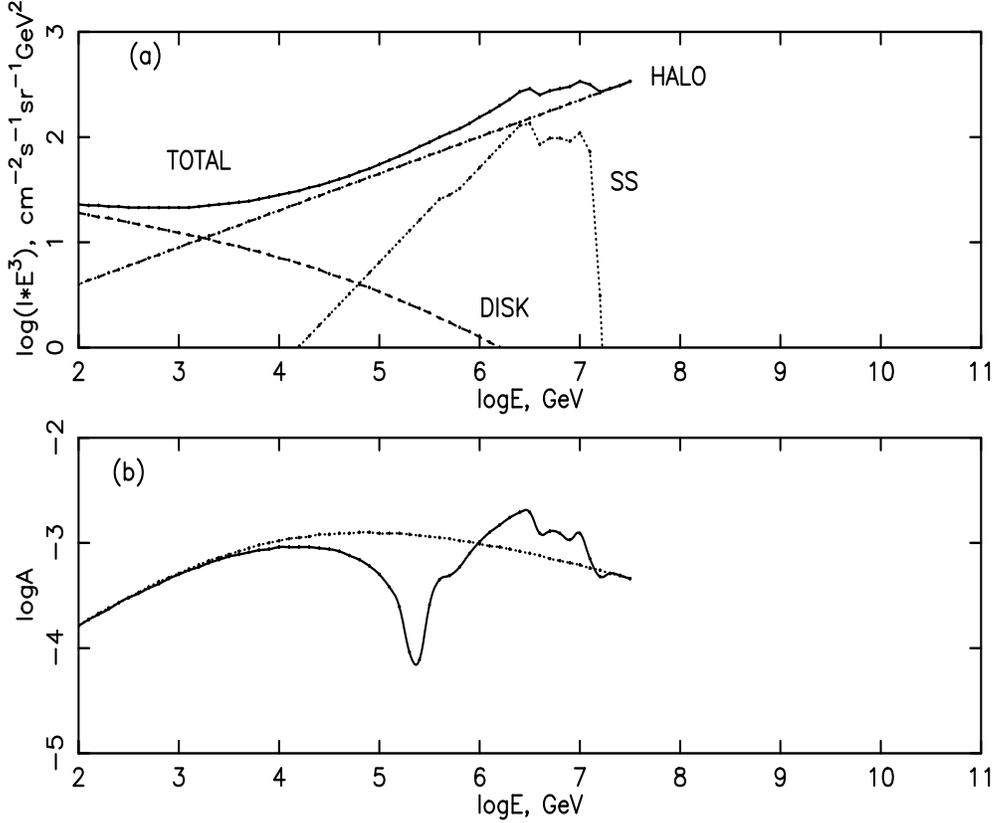}
\end{center}
\caption{\footnotesize (a) Schematic formation of the CR energy spectrum from the Disk,
 the Halo and the Single Source. 
(b) The amplitude of the first harmonic of the CR anisotropy obtained with 
contributions from Disk + Halo + Single Source (~full line~) and the same with 
contributions only from Disk + Halo without Single Source (~dotted line~).}   
\label{fig:fig2}
\end{figure}   

As for the Disk, we consider it as a cylinder with radius $R_d = 15 kpc$ and 
half thickness $H_d = 0.15 kpc$, therefore its volume is equal to 
$V_d = 2\pi H_d R_d^2 = 212 kpc^3$. Since we assume the dominance of the CR from the 
Halo we have to admit that it is difficult to extract the Disk CR from the total 
observed CR and our knowledge of their characteristics is poor. In order to make the 
contribution of the Disk to the observed CR intensity at sub-PeV energies small, we 
adopt  $\delta_d = 1.0$. The absolute normalisation is taken such that at lower, TeV 
and sub-TeV, energies CR from the Disk overcome CR from the Halo. The adopted spectrum 
from the Disk is also shown in Figure 2a and denoted as 'DISK'. The sum of Disk and 
Halo CR is denoted as 'TOTAL'.   
\section{Anisotropy}
We assume that the anisotropy appears only in the vicinity of the sources, i.e. in the
 Disk. The amplitude of the anisotropy $A$ is connected with the CR intensity $I$, its 
gradient, $gradI$, and the diffusion coefficient $D$ as $A=\frac{3DgradI}{cI}$. If the 
relative gradient $\frac{gradI}{I} \neq 0$ and $D \propto E^{\delta_d}$ then $A$ in the
 Disk rises with energy as $A_d \propto E^{\delta_d}$. The anisotropy of CR in the Halo
 is postulated as being $A_h = 0$ and the isotropy of CR re-entrant from the Halo back 
into the Disk dilutes the anisotropy of CR produced and trapped in the Disk.

In this treatment we just consider the first harmonic. Later work will deal with higher
 multipoles. We calculate the amplitude of the first harmonic for the case where only 
Disk and Halo contribute to CR as 
\begin{equation}
A = \frac{A_dI_d}{I_d+I_h}
\end{equation}
The result is shown in Figure 2b by the dotted line. The rise of the amplitude at 
energies above 100 GeV is due to the rise of the diffusion coefficient in the 
expression for $A_d$ mentioned above. The slow decrease of $A$ above $\sim$10 TeV is 
explained by the rising fraction of isotropic CR from the Halo, which overcomes the rise of $A_d$.
However, this scenario does not reproduce the remarkable dip in the amplitude visible 
in the experimental data at $logE,GeV = 5.3-5.5$ and the subsequent rise of the 
amplitude above this dip (~Figure 1a~).

We think that these features, if they are real, are connected with the existence 
of the Single Source, from which the CR energy spectrum is schematically shown by the 
dotted line in Figure 2a and denoted as 'SS'. However, this idea alone is not enough to
 reproduce the experimental data and here the examination of the phase of the first 
harmonic could help. In Figure 2b it is seen that after the moderate decrease in the 
energy interval 0.1 - 100 TeV the phase suddenly changes to its opposite. We consider 
this change seriously and propose that CR from the Single Source have a phase opposite 
to that of the background at lower energies. This is a necessary complementary 
requirement in our model. The rising part of CR coming from the opposite direction 
would reduce the anisotropy of the background from the Disk and Halo as
\begin{equation}
A = abs(\frac{A_dI_d-A_{ss}I_{ss}}{I_d+I_h+I_{ss}})
\end{equation}
The result of the calculations with the contribution of the Single Source is shown in 
Figure 2b by the full line. A comparison with the experimental data is shown also in 
Figure 1a by the full line. It is seen that after minimum in the dip the amplitude 
of the anisotropy starts rising again and it is caused by the rising contribution of 
the Single Source, which has the opposite phase.
 
Beyond the knee, and after the end of the contribution of the Single Source at about 
$10^8$GeV, the phase should return to its previous value. New measurements at these 
highest energies shown in Figure 1a \cite{PAO,Yakutsk2} demonstrate the rise of the 
anisotropy amplitude, but the comparison with the simulations led the PAO collaboration
to make a conservative conclusion that 'no clear evidence for anisotropy has been found
 yet' \cite{PAO}. We have to wait for results of higher statistical accuracy.
\section{Discussion}
We think that the described model with three basic ingredients: Disk, Halo and Single 
Source is reasonable and is worthy of discussion. The new features advocated here are: 

(a) The dominance of the Halo component in the sub-PeV region. It means that the CR 
which we observe and study in spite of being ourselves inside the Disk come mostly from
 the Halo. It is disputable but helps to understand the low anisotropy, small radial 
gradient of CR intensity and small level of irregularities in the regular power law 
energy spectrum. 

(b) The idea about the Single Source, which has to be nearby and young and creates the 
knee, usually raises questions: 'if it is nearby why we do not see it in the 
anisotropy ?'. It is a very reasonable question and this work gives the answer. The 
Single Source causes the stronger decrease and the dip in the amplitude of the dipole 
anisotropy
at sub-PeV energies. It is also seen in the change of the phase of the anisotropy at 
the same energies. It means that the Single Source should deliver CR from the direction
opposite to the direction of CR from the background and it is a new assumption in the 
Single Source scenario. Above the dip energy the amplitude starts to rise again with 
the opposite phase, as expected.

The dip in this model appears as the result of subtraction of two 
bigger values. Its position and shape are extremely sensitive to the choice of
parameters participating in expression (4). The relatively good agreement with the 
experimental data is the result of the fitting procedure, but, nevertheless: \\
(i) it demonstrates the possibility of achieving agreement within the 
framework of our simplistic model and \\
(ii) the high sensitivity of the dip to the input parameters of the expression (4) 
gives the possibility of investigating these parameters when precise results in this 
energy region are obtained.         

(iii) The phase of the first harmonic in the PeV region, where the contribution of the
Single Source is big enough, could help to locate it on the sky. The present 
experimental data have a too big spread to make a conclusion.  
   
We understand that this scenario raises more questions than gives the answers. For 
instance, the main questions are:

(a) Do the Halo and the Single Source really exist ? Arguments for positive answers
are given in the \S3, but more supportive arguments are needed.

(b) Why numerical estimates for CR intensity in the Disk and the Halo, calculated
with the expression (2), are smaller than in the observations~?
    
(c) Why the energy spectrum of CR in the Disk is steeper than the spectrum in the Halo 
~? According to our conception developed in \cite{EW1} the spectrum in the Disk, with 
its higher turbulence in the interstellar medium due to SN explosions, should be flatter
 than in the Halo where there are no such powerful sources of turbulence as SN.

(d) To what extent are the simplified assumptions about the shapes and normalisation of
 the Disk, Halo and Single Source spectra as well as other parameters: $A_d, A_h$ and 
$A_{ss}$, reasonable and what the more sophisticated approach will do for the result ?

The answers to these and other questions are the subject of further work.
\section{Conclusions}
We develop a model which helps to give an explanation  for at least one of the three 
puzzles mentioned in the Introduction, viz. the small radial gradient of the CR 
intensity, the small magnitude and peculiar energy dependence of the CR anisotropy and 
the small level of irregularities in the CR energy spectrum. In this paper we analyse 
the second puzzle - the anisotropy. 

The model exploits three basic ingradients: Disk, Halo and the Single Source. We 
postulate the dominant role of the Halo and its CR in our observations in spite of the 
fact that we
are located inside the Disk. At PeV energies, approaching the knee, contributions from
the Single Source begins to play an important role and inspired by the experimental
evidence we assume that the phase of the CR intensity from the Single Source is 
opposite to the phase of the background CR from the Disk and the Halo. Due to this 
effect the amplitude of the dipole anisotropy decreases and approaches a minimum 
(dip) at sub-PeV energies. After that the amplitude begins to rise again, but because
CR are mainly from the Single Source they come preferentially from the direction
opposite to that at lower sub-PeV energies. The position and the shape of the dip is 
extremely sensitive to the parameters of the spectra adopted for the three ingredients:
 Disk, Halo and the Single Source. This sensitivity can be used for the study of the CR
 origin in the vicinity of the knee in the PeV energy region. \\

{\bf Acknowledgements} 

The authors are grateful to the Kohn Foundation for financial support. The (unknown)
reviewer is thanked for very helpful comments and suggestions.

\end{document}